\documentstyle[epsfig]{mn}
\begin{document}

\title[GRS~1915+105]
{Spectral, polarisation and time--lag properties of GRS 1915+105
radio oscillations}
\author[R.~P.~Fender et al.]
{R. P. Fender$^1$
D. Rayner$^2$, 
S. A. Trushkin$^3$,  K. O'Brien$^1$,
R. J. Sault$^2$, 
\cr G. G. Pooley$^4$, R.P. Norris$^2$
\\
$^1$ Astronomical Institute `Anton Pannekoek', University of Amsterdam,
and Center for High Energy Astrophysics, Kruislaan 403, \\
1098 SJ, Amsterdam, The Netherlands {\bf rpf@astro.uva.nl}\\
$^2$ Australia Telescope National Facility, CSIRO, PO Box 76, Epping
NSW 2121, Australia\\
$^3$ Special Astrophysical Observatory, RAS, Nizhnij Arkhyz,
Karachaevo-Cherkassia 369167, Russia\\ 
$^4$ Mullard Radio Astronomy Observatory, Cavendish Laboratory,
Madingley Road, Cambridge CB3 OHE {\bf ggp1@cam.ac.uk}\\
}

\maketitle

\begin{abstract}

We report high sensitivity dual-frequency observations of radio
oscillations from GRS 1915+105 following the decay of a major flare
event in 2000 July. The oscillations are clearly observed at both
frequencies, and the time-resolved spectral index traces the events
between optically thin and thick states.  While previously anticipated
from sparse observations and simple theory, this is the first time a
quasi-periodic signal has been seen in the radio spectrum, and is a
clear demonstration that flat radio spectra can arise from the
combination of emission from optically thick and thin regions.  In
addition, we measure the linear polarisation of the oscillations, at
both frequencies, at a level of about 1--2\%, with a flat
spectrum. Cross-correlating the two light curves we find a mean delay,
in the sense that the emission at 8640 MHz leads that at 4800 MHz, of
around 600 seconds. Comparison with frequency-dependent time delays
reported in the literature reveals that this delay is variable between
epochs. We briefly discuss possible origins for a varying time delay,
and suggest possible consequences.

\end{abstract}

\begin{keywords}

binaries: close -- stars: individual: GRS~1915+105 -- infrared: stars
-- radio continuum: stars -- ISM:jets and outflows

\end{keywords}


\begin{figure}
\centering
\leavevmode\epsfig{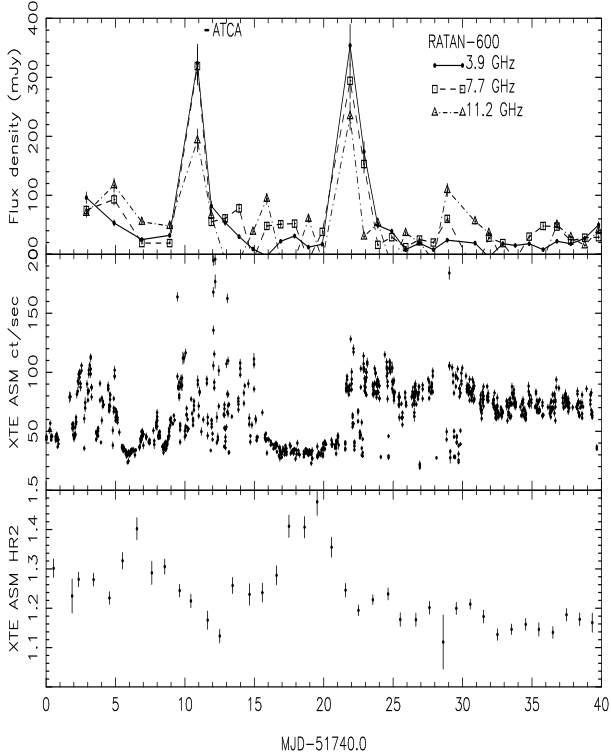}
\caption{Radio and X-ray monitoring of GRS 1915+105 before and after
the ATCA observations (which are indicated by a bar in the top panel).
Note that the XTE ASM total count rate is based upon individual scans,
whereas the HR2 colour is based on daily averages.
The RATAN radio data reveal two major flaring episodes. The first,
during the decay of which our ATCA observations were made, may be
associated with the brief hard X-ray state around MJD 51746, or may be
associated with the subsequent week-long `plateau' between MJD 51755--62.}
\label{}
\end{figure}

\section{Introduction}

GRS 1915+105 is a luminous and spectacularly variable source of
radiation from radio through to hard X-ray regimes. Its behaviour in
both hard and soft X-rays is unique (e.g. Foster et al. 1996; Morgan,
Remillard \& Greiner 1997; Belloni et al. 2000), and it is a celebrated
source of relativistic jets (Mirabel \& Rodr\'\i guez 1994; Fender et
al. 1999; Rodr\'\i guez \& Mirabel 1999; Dhawan,
Mirabel \& Rodr\'\i guez 2000, Giovannini et al. 2001).

X-ray flux variations on a variety of timescales have been interpreted
as the repeated draining and refilling of the inner accretion disc,
possibly due to extremely rapid transitions between `canonical' black
hole accretion states (e.g. Belloni et al. 1997a,b; Feroci et
al. 1999; Belloni et al. 2000); however the physical meaning of these
apparent changes in inner disc radius is not entirely certain
(Merloni, Fabian \& Ross 2000).  Quasi-sinusoidal oscillations with
similar periods, almost certainly the signature of synchrotron
emission from repeated ejection events, have been observed at radio,
millimetre and infrared wavelengths (Pooley \& Fender 1997; Fender et
al. 1997; Eikenberry et al. 1998, 2000; Mirabel et al. 1998; Fender \&
Pooley 1998, 2000). These oscillations appear to have a direct
connection to the X-ray dips (Pooley \& Fender 1997; Eikenberry et
al. 1998, Mirabel et al. 1998, Klein-Wolt et al. 2001), although there
is some debate as to whether they are associated with `soft'
(e.g. Naik \& Rao 2000) or `hard' (e.g. Klein-Wolt et al. 2001) X-ray
states.  Delays between different radio bands (Pooley \& Fender 1997;
Mirabel et al. 1998) and between the infrared and radio bands (Mirabel
et al. 1998; Fender \& Pooley 1998) clearly indicate that optical
depth effects play an important role in the observed emission from
these ejections. Details such as the magnitude and variability of
these delays are crucial for gaining a quantitative understanding of
the outflow process.

\section{Observations}

\subsection{RATAN-600}

The RATAN-600 observations were carried out as a part of a monitoring
program of microquasars to study their flare activity across a broad
frequency range. Observations were performed at 3.9, 7.7, 11.2 and
21.7 GHz; the 3--11 GHz data are presented in the top panel of Fig 1.
The flux density calibration was performed using observations of 3C286
(1328+30) and PKS1345+12.  Although interference sometimes prevented
realisation of the maximum sensitivity of the radiometers, daily
observations of reference sources indicate that the error in the flux
density measurements for 1915+10 did not exceed 5-10\% at 2.3, 3.9,
and 11.2 GHz and 10-15\% at 21.7 GHz. For further details see
Trushkin, Majorova \& Bursov (2001).

\subsection{XTE ASM}

We have made use of public data from the Rossi X-ray Timing Experiment
(XTE) All-Sky Monitor (ASM; Levine et al. 1996). These data are
available at {\bf xte.mit.edu}. The total intensity and HR2 X-ray
colour (ratio of counts in 5--12 to 3--5 keV bands) are presented in
Fig 1.

\subsection{ATCA}
We have observed GRS 1915+105 with the Australia Telescope Compact
Array (ATCA) for approximately six hours on 2000 July 26. The override
observations were triggered as a result of the radio flare observed by
the RATAN-600 telescope around MJD 51750.  At this time the array was
in a compact configuration. As a result, in order to reduce the
effects of field sources (see e.g. Chaty et al. 2001 for radio images
of the field) we only used interferometer baselines $\geq 2200$m at
4800 MHz. The compact configuration and relatively poor hour-angle
coverage precluded any attempt to confidently calibrate Stokes V, and
so we were not able to make a circular polarisation measurement.

In Fig 2 we present dual-frequency total intensity, spectral index and
linear polarisation measurements of GRS 1915+105 obtained during the
ATCA run. Care has been taken to check the reality of the low-level
linear polarisation (ie. Stokes Q, U) measurements, and we are
confident that those presented in Fig 2 are realistic. Mapping the
data in linear polarisation produces results consistent with averaging
the data presented in Fig 2. Note however that for the last few scans
the polarisation calibration solutions were not good and those data
are not used.

\section{Results}

The 40-day light curves presented in Fig 1 clearly reveal the
existence of radio flaring and dramatically varying X-ray flux and
hardness. It has been previously established (e.g. Foster et al. 1996,
Fender et al. 1999) that the prolonged (ie. more than a few days) hard
states, or `plateaux', are accompanied by flat or inverted spectrum
radio emission and appear to be always followed by an optically thin
radio flare. The inverted-spectrum emission probably corresponds to a
powerful self-absorbed quasi-continuous jet (Dhawan et al. 2000;
Fender 2001) and the optically thin post-plateau flares have been
directly resolved into relativistically outflowing components
(e.g. Mirabel \& Rodriguez 1994; Fender et al. 1999). In terms of Fig
1, we consider the plateau phase to be between MJD 51756--51762, and
its resultant flare to be the radio event which peaked around MJD
51762.  What is less clear is whether the flare of $\sim$ MJD 51750.0
was related to this subsequent plateau or to the shorter X-ray hard
state around MJD 51746--51747.  It is the decay of this flare, and the
subsequent emergence of core-oscillation events, which were studied in
detail with our ATCA observations. Note that there appears to be a
third smaller flare associated with a drop in the X-ray flux around
MJD 51768.

\begin{figure}
\centering
\leavevmode\epsfig{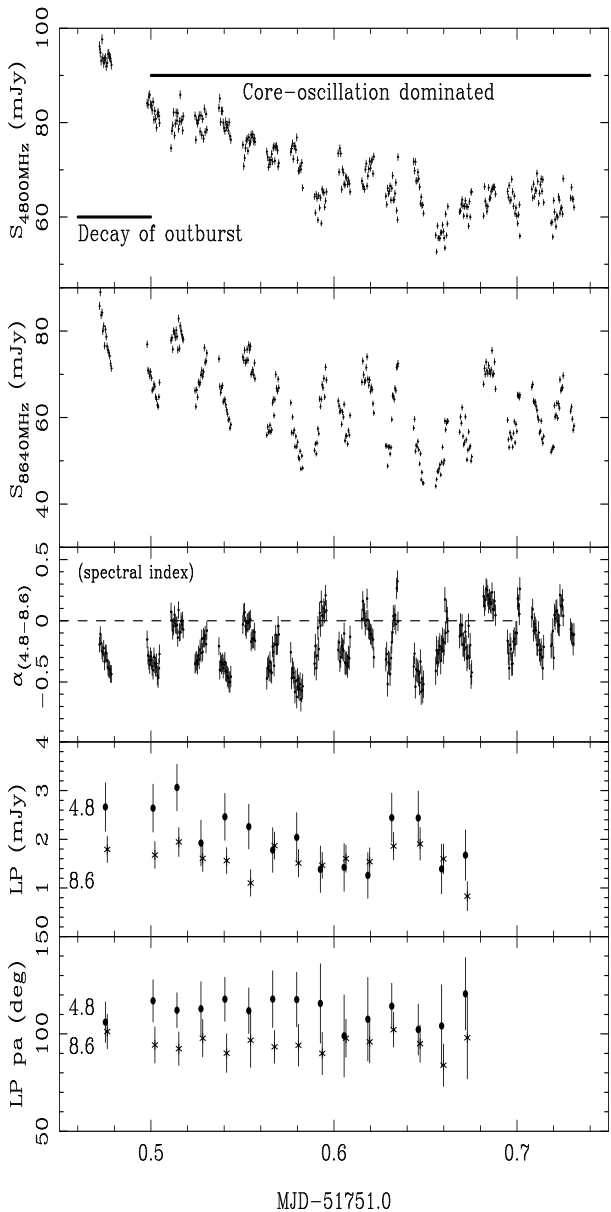}
\caption{Dual-frequency observations of the transition from outburst
to oscillating states in the radio emission from GRS 1915+105,
following the first major flare indicated in Fig 1.  The top three
panels present 30-sec averaged data, the lower two panels 10-min
averaged data.  The spectral index, in the middle panel, is defined 
as $\alpha = \Delta \log S_{\nu} / \Delta \log \nu$.
After $\sim$MJD 51751.55 the radio variability is dominated
by oscillations, with a very clear signature in the spectral
index. Linear polarisation, at the level of a few \%, is clearly
detected throughout the observation; note however that the
polarisation calibration solutions became erratic for the last few
scans and so those data are not presented here.
}
\label{}
\end{figure}

\subsection{Optically thin decay}

The characteristic decay and optically thin spectral index during the
first two scans with ATCA (MJD 51751.45--51751.50) are indicative of
ejecta from the observed outburst fading away as they propagate
through the ISM, as directly imaged for previous outbursts by
e.g. Mirabel \& Rodriguez (1994), Fender et al. (1999).  At a bulk
velocity of $\geq 0.9c$ these synchrotron emitting clouds are likely
to be by this stage hundreds of AU from the system and totally
decoupled from the accretion process.

During these period the mean total intensities observed from GRS
1915+105 were $\sim 90$ mJy (4.8 GHz), and $\sim 70$ mJy (8.64 GHz),
corresponding to a spectral index $\alpha = \Delta\log S_{\nu} /
\Delta \log \nu \sim -0.4$.

\subsection{Core oscillation events}

\begin{figure}
\centering
\leavevmode\epsfig{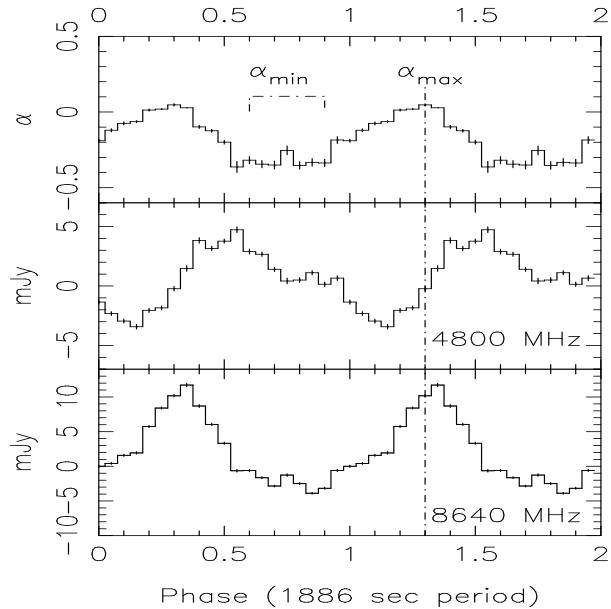}
\caption{Spectral index, total flux density at 4800 and 8640 MHz,
folded on a best-fit period for the oscillations of 0.02144d = 30.87
minutes, in 20 equally-sized phase bins.
Data from the first two scans (ie. prior to MJD 51751.5) have
been excluded as they were dominated by the outburst decay. 
}
\label{}
\end{figure}


After MJD 51751.51, the light curve and spectral index indicate a
transition to radio oscillation events (Pooley \& Fender 1997; Fender
\& Pooley 1998, 2000) which are known to be associated with rapid
changes in the accretion flow (as indicated by simultaneous radio and
X-ray observations -- Pooley \& Fender 1997; Mirabel et al. 1998;
Klein-Wolt et al. 2001).

After this point the spectral index as plotted in Fig 2 clearly shows
oscillations between optically thin ($\alpha_{\rm min} \sim -0.7$) and
thick ($\alpha_{\rm max} \sim +0.2$). While it could be qualitatively
appreciated from previous dual-frequency radio observations
(e.g. Pooley \& Fender 1997; Mirabel et al. 1998), this is by far the
clearest illustration to date of the rapid changes in the spectral
index of the emission from GRS 1915+105 during the radio oscillation
events.


\subsection{Polarisation and rotation measure}

Linear polarisation was measured at both frequencies from GRS 1915+105
during both the optically thin and core-oscillation phases of our
observations. 

During the optically thin phase the mean linearly polarised flux
densities were $\sim 3$ mJy (4.8 GHz) and $\sim 1$ mJy (8.64 GHz),
corresponding to fractional linear polarisations were $\sim 3$\% (4.8
GHz) and $\sim 2$\% (8.6 GHz). During the transition to the
oscillation phase the polarised flux density at 4.8 GHz drops
significantly and becomes comparable to that at 8.6 GHz as the total
intensity spectrum similarly flattens. The flattening of the linear
polarisation spectrum, as well as the apparent halt in its initial
decay at 4800 MHz, strongly suggest that it is associated with the
core events and not remnant emission from the major optically thin
ejection.  Therefore we conclude that our data measure around 1 mJy of
linearly polarised flux density associated with the oscillations at
each frequency, corresponding to a fractional linear polarisation of
1--2\%.

The position angles of the linear polarisation are similar for both
phases, with averages for the entire run of pa(4800 MHz) $=(112 \pm
2)^{\circ}$ and pa(8640 MHz) $=(96 \pm 2)^{\circ}$. The 4800 MHz
position angle is similar to that measured by MERLIN for the inner
regions of the jet in Fender et al. (1999). Ascribing the difference
in position angles to Faraday rotation, we can estimate a minimum
rotation meausure of $\sim 100$ rad m$^{-2}$, with an associated
intrinsic position angle of $\sim 90^{\circ}$ (neither exactly
parallel or perpendicular to the observed jet axis). This is
consistent with rotation measures of radio pulsars at distances of
several kpc or more (e.g. Han, Manchester \& Qiao 1999).  However,
this rotation measure seems to contradict the observations of Rodr\'\i
guez et al. (1995) whose measurements of linear polarisation at three
frequencies seem to imply an upper limit to the rotation measure of
$\sim 50$ rad m$^{-2}$.  Possibly the magnetic field structure in the
jet rotates between the regions from which the 8640 and (downstream)
4800 MHz emission arise.  Whatever the intrinsic polarisation angle,
the fact that the observed position angles remain approximately
constant indicates the persistence over the observation of some
magnetic field structure.

\subsection{`Pulse folding' the data}

In order to understand the oscillations better, we have attempted to
fold the data onto a single period. Using the Starlink package PERIOD,
we have determined a best-fit single period of 0.02144 days, and
folded the data onto that period. The detrended (see below)
folded spectral index and total
intensity light curves are plotted in Fig 3. The technique worked
surprisingly well -- there is a clear modulation in the folded light
curves at both frequencies and in the spectral index, $\alpha$. In
fact the same period in the signal, to within 2\%, was found in
independent period searches on the 4800 and 8640 MHz data.  It is
clear that variations in $\alpha$ are dominated by the 8640 MHz light
curve, which has significantly larger degree of modulation and as a
result a clearer profile. From Fig 3 we can see that the lag between
the two light curves is about 0.25--0.35 in phase, corresponding to 7
-- 11 minutes, consistent with the lag obtained in the
cross-correlation analysis below.
We have also attempted to fold the polarised flux and position angle
on the oscillation period. There is a hint in the folded light curves
that the linearly polarised flux is modulating, but it is not
statistically significant. 

\section{Optical depth time delays}

We have performed a cross-correlation analysis of the two lightcurves
shown in Figure 2, using the discrete correlation function (DCF) of
Edelson \& Krolik (1988). Unlike the Interpolated Correlation Function
(ICF) of Gaskell \& Peterson (1987), which interpolates between
individual points of the driving lightcurve (the lightcurve at 8640
MHz), the DCF determines the correlation coefficient at discrete time
lags, which are then binned to create a uniformly sampled
cross-correlation function. The DCF was preferred due to the large,
quasi-regular gaps in the data and its ability to attach meaningful
error-bars to the resulting correlation function.

In order to remove the long term trend from the lightcurves, which
could introduce spurious correlations into the correlation function,
we subtracted a quadratic fit to the two data sets (`detrending')
before calculating
the DCF. The resulting DCF is shown in Figure 4; the strongest peak of the
correlation function has a lag of 625 seconds, with a correlation
coefficient of 0.47$\pm$0.08, although the feature is broad, which may
indicate that a range of lags are present and/or reflect the less
distinct pulse shape at the lower frequency (Fig 3). The DCF also
shows secondary peaks at approximately $\pm$ 1800 seconds from the
main peak, which are present in the auto-correlation function and are
due to the periodic nature of the variability.

\begin{figure}
\begin{center}
\epsfig{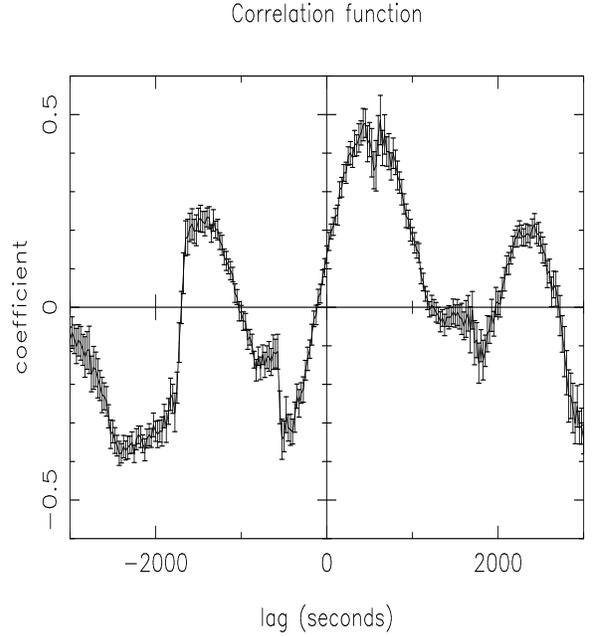}
\end{center}
\label{dcfplot}
\caption{The discrete correlation function for the lightcurves at 8640 and
4800 MHz. A positive lag is defined to occur when
variations in the 8640 MHz lightcurve
precede those in the 4800MHz lightcurve.}
\end{figure}
 
\begin{table}
\caption{Summary of frequency-dependent time delays in GRS 1915+105}
\begin{tabular}{ccccc}
\hline
MJD & $\nu_1$ (Hz) & $\nu_2$ (Hz) & $\Delta t$ (sec) & \\
\hline
50227 & $15 \times 10^{9}$ & $8.3 \times 10^{9}$ & $\sim 250$ & PF97 \\
\hline
50583 & $15 \times 10^{9}$ & $8.3 \times 10^{9}$ & $\sim 750$ & M98 \\
      & $8.3 \times 10^{9}$ & $5 \times 10^{9}$ & $\sim 940$ & M98 \\ 
\hline
50700 & $1.4 \times 10^{14}$ & $8.3 \times 10^{9}$ & $\sim 860$ & M98 \\
50705 & $1.4 \times 10^{14}$ & $1.5 \times 10^{10}$ & $\sim 420$ /
$\sim 2000$ & FP98 \\
51318 & $1.4 \times 10^{14}$ & $1.5 \times 10^{11}$ & $\leq 300$ &
FP00 \\
51751 & $8.6 \times 10^{9}$ & $4.8 \times 10^{9}$ & $\sim 600$ & F01 \\
\hline 
\end{tabular}
\noindent
PF97 = Pooley \& Fender 1997; M98 = Mirabel et al. 1998;
FP98, FP00 = Fender \& Pooley 1998, 2000; F01 = this paper. 
The data from MJD 50583 (M98) were recorded simultaneously at {\em
three} radio wavelengths.
\end{table}

In Table 1 we summarise the frequency-dependent time lags reported for
the radio--infrared oscillations in GRS 1915+105. It is clearly of
interest to know whether these times lags are variable.  From table 1
it is clear that the data reported here are now the {\em second} case
of different delays between the same two frequencies:

\begin{enumerate}
	\item{Pooley \& Fender (1997) reported delays between 15 GHz
and 8.3 GHz which were significantly shorter than those reported by
Mirabel et al. (1998).}
	\item{The delay reported by Mirabel et al. (1998) between 5 --
8 GHz was also considerably longer than that reported between almost
the same frequencies in this paper.}
\end{enumerate}

What could cause different time delays at different epochs ? There are
two obvious possibilities -- different physical size scales at
different epochs, or changes in the Doppler factor of the jet (since
the delay between different frequencies originates in the relativistically
flowing medium). In reality there may be some combination of these,
and other, effects.

If the first explanation is the main reason however, we may expect some
observational relation between radio flux levels and delays. The
simultaneous 8 \& 15 GHz light curves presented in Pooley \& Fender
(1997) had flux densities in the range 50--100 mJy. The oscillations
presented in this paper have a range of about 40--100 mJy. The
amplitudes reported in Mirabel et al. (1998), up to 100 mJy, appear to
be somewhat larger. Thus the longer delays may be due to jet structure
which is larger overall, and we would expect, given more data, to see a
correlation between delay length and oscillation amplitude.

If the explanation is instead due to a changing Doppler factor, this
may be most naturally explained as being due to a change in the angle
of the jet to the line of sight (although we cannot rule out changes
in the bulk Lorentz factor), for example if the jet is precessing. For
a bulk Lorentz factor of 5 and angle of jet to the line of sight of
$\sim 65$ degrees, as calculated for GRS 1915+105 ejections for a
distance of 11 kpc (Fender et al. 1999), a swing of $\ga 20$ degrees
would be required in order to change the Doppler factor by the factor
of $\sim 2$ indicated by table 1 (for larger Lorentz factors a smaller
swing in angle is required, but this would have further implications
for e.g. energetics). In the event that varying delays were caused by
a periodically varying jet angle to the line of sight, we might expect
to see, in the long-term, a repeating pattern of `long' and `short'
delays.

\section{Conclusions}


In a set of high-sensitivity dual-frequency radio observations of GRS
1915+105 with ATCA during a period of radio oscillations we have
captured in detail some key characteristics. 

Firstly we have clearly shown that the GHz radio spectrum of GRS
1915+105 repeatedly cycles between optically thick and thin during
periods of radio oscillations. While this could be qualitatively
appreciated from previous sparse observations and simple concepts, we
have for the first time captured the strong quasi-periodic signal in
the radio spectrum. It can now be directly appreciated, from Fig 2,
that were we to average over multiple oscillation events (for e.g. a
weaker source or a shorter oscillation period), we would measure a
spectral index which was a combination of both optically thick and
optically thin emission, with an average close to zero, as seen in
hard states of black hole candidate X-ray binaries (Fender 2001).  The
transition from optically thin to flat/inverted radio spectra
following the outburst is also reminiscent of the hard state
transients (see e.g. Fig 3 of Fender 2001).  In addition, the $\sim
1$\% level of linear polarisation associated with the oscillations at
both frequencies is comparable to that measured for the flat spectrum
in the hard states of two black hole candidates, GS 2033+338/V404 Cyg
(Han \& Hjellming 1992), and GX 339-4 (Corbel et al. 2000), and this
level of linear polarisation may be a generic feature of such flat
spectra in BHC low/hard states (Fender 2001).  The similarity of the
linear polarisation position angle for both the optically thin flare
and oscillations may indicate a preferred orientation of magnetic
field for the outflow.  In addition in this case we have measured the
spectrum of the linear polarisation, and find it also to be
approximately flat.


Finally, we have performed a cross-correlation analysis of the two
radio lightcurves in order to establish a mean delay. This is the most
thorough analysis of the time delays to date; previous attempts were
`by eye' or involved only the peaks of single events. Comparing all
the reported frequency-dependent time delays for the oscillations from
the literature, we find that there are significant differences at
different epochs. The longest time delays reported seem to be
associated with the large-amplitude oscillations in Mirabel et
al. (1998), which may indicate a relation between event amplitude and
subsequent time delay. However, at this stage a changing Doppler
factor or some other effect cannot be ruled out.

Detailed observations such as these are crucial to a quantitative
understanding of the physics involved in jet production, and may be
vital in distinguishing between, for example, models of discrete
ejections (e.g. van der Laan 1966; Mirabel et al. 1998)
or shocks propagating along quasi-continuous flows (e.g. Blandford \&
K\"onigl 1979; Kaiser, Sunyaev \& Spruit 2000).

\section*{Acknowledgements}

We would like to thank the referee, Ralph Spencer, for constructive
criticism of this paper, and R. Ramachandran for a discussion about radio
pulsar rotation measures.  S.A.T. is thankful to RFBR for support
(grant N98-02-17577).  The Australia Telescope is funded by the
Commonwealth of Australia for operation as a National Facility managed
by CSIRO.  RXTE ASM results were provided by the ASM/RXTE teams at MIT
and at the RXTE SOF and GOF at NASA's GSFC.

\end{document}